\documentclass[aps,preprint,preprintnumber,nofootinbib,amsmath,amssymb,bm,12pt]{revtex4}
\usepackage{graphicx,color}
\usepackage{bm,amsmath,amssymb}
\usepackage{ascmac}
\allowdisplaybreaks

\begin{document}

{
\small
\rightline{
\baselineskip16pt\rm\vbox to20pt{
\hbox{OCU-PHYS-525}
\hbox{AP-GR-162}
} } } 

\author{Ken~Matsuno}
\email{matsuno@sci.osaka-cu.ac.jp}

\affiliation{
Department of Mathematics and Physics, Graduate School of Science, Osaka City University,
Sumiyoshi, Osaka 558-8585, Japan
\bigskip}

\vskip 1cm

\title{ {\Large 
Light deflection by squashed Kaluza-Klein black holes in a plasma medium 
} \\ }

\begin{abstract}
We study motions of photons in an unmagnetized cold homogeneous plasma medium 
in the five-dimensional charged static squashed Kaluza-Klein black hole spacetime.  
In this case, a photon behaves as a massive particle 
in a four-dimensional spherically symmetric spacetime.   
We consider the light deflection by the squashed Kaluza-Klein black hole 
surrounded by the plasma in a weak-field limit.   
We derive corrections of the deflection angle to general relativity,  
which are related to the size of the extra dimension, the charge of the black hole 
and the ratio between the plasma and the photon frequencies.     
\end{abstract}

\maketitle

\section{Introduction}
\label{sec:intro}

The direct detection of gravitational waves generated by the coalescence of black hole binaries 
\cite{Abbott:2016blz} 
and the successful imaging of immediate vicinity of a supermassive black hole candidate 
in the center of the galaxy M87 \cite{Akiyama:2019cqa} 
mean that researches of black holes have entered a new stage.  
Motivated by these astrophysical observations, 
we are interested in performing in-depth studies of 
the optical features of higher-dimensional black hole solutions.  
Recently, verifications of extra dimensions and braneworld black holes 
by observations of black hole shadow have been studied 
\cite{Vagnozzi:2019apd, Banerjee:2019nnj, Nucamendi:2019qsn, Neves:2020doc}.   
In this paper, we focus on the light deflection by higher-dimensional black holes 
surrounded by a plasma in the spacetime with compact extra dimensions.

Higher-dimensional black hole spacetimes are actively discussed  
in the context of string theories and braneworld models. 
If higher-dimensional black hole solutions have compactified extra dimensions, 
we can regard such black hole solutions as candidates of realistic models, 
since our observable world is effectively four dimensional. 
We call these Kaluza-Klein black holes. 
In four-dimensional general relativity, the gravitational field in vacuum with spherical symmetry 
is uniquely described by the Schwarzschild metric. 
However, in a higher-dimensional spacetime with Kaluza-Klein structure, 
even if we impose asymptotic flatness in a four-dimensional section, 
the metric is not determined uniquely. 
A family of five-dimensional squashed Kaluza-Klein black hole solutions 
\cite{Dobiasch:1981vh, Gibbons:1985ac, Ishihara:2005dp, Stelea:2008tt, Allahverdizadeh:2009ay, Tomizawa:2012nk} 
asymptote to effective four-dimensional spacetimes with a twisted S$^1$ 
as an extra dimension at infinity  
and represent fully five-dimensional black holes near the squashed S$^3$ horizons.  
Then squashed Kaluza-Klein black hole solutions with a twisted compactified extra dimension 
would describe the geometry around the compact objects.  
Several aspects of squashed Kaluza-Klein black holes are discussed, for example, 
multiblack holes \cite{Matsuno:2008fn, Chen:2010ih, Matsuno:2015ega},  
stabilities \cite{Kimura:2007cr, Kimura:2018whv}, 
quasinormal modes \cite{Ishihara:2008re, He:2008im, He:2008kq}, 
thin accretion disk \cite{Chen:2011wb},  
x-ray reflection spectroscopy \cite{Zhu:2020cfn}, 
gyroscope precession \cite{Matsuno:2009nz, Azreg-Ainou:2019ylk}, 
strong gravitational lensing \cite{Liu:2010wh, Chen:2011ef, Sadeghi:2012bj, Sadeghi:2013ssa, Ji:2013xua} 
and black hole shadow \cite{Long:2019nox, 1828181}.

The gravitational lensing combines a wide range of phenomena 
connected with the deflection of light rays by a gravitational field.     
Most gravitational lensing deals with geometrical optics in vacuum 
and uses a notion of the deflection angle \cite{Virbhadra:1999nm, Virbhadra:2008ws}.    
A basic assumption is the approximation of a small deflection angle of a photon   
which is well satisfied in some astrophysical situations related to the gravitational lensing.     
Since the photon trajectories and deflection angles of photons in vacuum 
do not depend on the photon frequency or its energy,  
the gravitational lensing in vacuum is achromatic.   
Then it is interesting to consider how the light trajectory and its deflection angle 
change in the presence of a plasma 
since light rays propagate through plasmas 
around compact objects, galaxies and galaxy clusters in the Universe.     
Self-consistent approaches for the geometrical optics in an arbitrary medium in a curved spacetime  
are discussed in the references \cite{Synge(1960), Perlick(2000)}.  
One of the most interesting effects of this kind is a chromatic gravitational deflection of light. 
In an unmagnetized cold homogeneous plasma medium,      
the refractive index of a plasma and the propagation of light rays depend on the photon frequency.    
Then the gravitational deflection of light is different from the vacuum case 
and its deflection angle depends on the ratio between the plasma and the photon frequencies 
\cite{BisnovatyiKogan:2008yg, BisnovatyiKogan:2010ar}.    
Since the effect of difference in gravitational deflection angles is significant 
for photons of longer wavelengths, 
the modifications of gravitational lensing due to the presence of the plasma 
is negligible in the visible spectrum.  
Then there are some astrophysical observations in the radio spectrum which would detect 
such plasma effects in low-frequency bands  
\cite{Muhleman:1966, Muhleman:1970zz, Lebach:1995zz, Shapiro:2004zz, 
Lonsdale:2009cb, vanHaarlem:2013dsa, Bentum:2016ekl, Turyshev:2018gjj}.  
Motivated by these observations,  
the influence of plasma media on the trajectory of light rays and the deflection angles of photons  
around compact objects have been studied in a variety of spacetimes  
including vacuum, electrovacuum and with a vast array of scalar fields or effective fluids 
at both finite and infinite distances in both weak and strong-field limits 
\cite{Tsupko:2013cqa, Morozova2013, Er:2013efa, Rogers:2015dla, Perlick:2015vta, 
Abdujabbarov:2016efm, Abdujabbarov:2015pqp, Rogers:2016xcc, Liu:2016eju, 
Perlick:2017fio, Abdujabbarov:2017pfw, Er:2017lue, Schee:2017hof, 
Crisnejo:2018uyn, Ovgun:2019wej, Crisnejo:2018ppm, Ovgun:2018oxk, Turimov:2018ttf, 
Crisnejo:2019xtp, Javed:2019rrg, Javed:2019ynm, Crisnejo:2019ril, Javed:2020fli, 
Ovgun:2020gjz, Javed:2020lsg, Javed:2020frq, Javed:2020wsv, Kumaran:2019qqp, 
ElMoumni:2020wrf, Fathi:2020otm, Babar:2020txt, Belhaj:2020rdb}.

In this paper, we investigate motions of photons and its deflection angles in a weak-field limit 
in the five-dimensional charged static squashed Kaluza-Klein black hole spacetime  
in the presence of a plasma.       
To the best our knowledge, photon motions around compact objects in plasma media   
have not been discussed in asymptotically Kaluza-Klein spacetimes.  
In the present work, we extend the derivations of weak deflection angles of photons 
in an unmagnetized cold homogeneous plasma medium in four-dimensional black hole spacetimes   
to the case of the five-dimensional squashed Kaluza-Klein black hole surrounded by such plasma.

This paper is organized as follows. 
In Sec.~\ref{sec:kkbh}, 
we review the properties of five-dimensional charged static Kaluza-Klein black hole solutions 
with squashed horizons. 
In Sec.~\ref{sec:photonmotions}, 
we consider photon motions in a homogeneous plasma medium in squashed Kaluza-Klein geometry 
and show that there is a stable circular orbit of a photon with no momentum 
in the direction of the extra dimension.  
In Sec.~\ref{sec:defangs}, 
we study the light deflection by the squashed Kaluza-Klein black hole 
surrounded by the homogeneous plasma in a weak-field limit.   
It is shown that the asymptotically Kaluza-Klein structure, the Maxwell field and the plasma  
modify the deflection angle of photon in the black hole geometry.   
Section \ref{sec:summary} is devoted to summary and discussion.

\section{Squashed Kaluza-Klein black holes}
\label{sec:kkbh}

We consider the charged static Kaluza-Klein black holes with squashed S$^3$ horizons, 
which are exact solutions of the five-dimensional Einstein-Maxwell theory \cite{Ishihara:2005dp}. 
The metric and the Maxwell field are given by 
\begin{align}
 & ds^2 = -F dt^2 + \frac{K^2 }{F} d\rho^2 + \rho^2 K^2 \left( d\theta^2 + \sin ^2 \theta d\phi^2 \right) 
 + \frac{r_\infty ^2}{4K^2} \left( d\psi + \cos \theta d\phi \right)^2 , 
\label{met}
\\
 & A_\mu dx^\mu = \frac{\sqrt 3 Q }{2 \rho} dt , 
\end{align}
with
\begin{align}
 F = 1 - \frac{2 M }{\rho } + \frac{Q^2 }{\rho^2 } , \quad 
 K^2 = 1 + \frac{\rho_0}{\rho }  ,
\end{align}
where the parameters $M ,~ Q ,~ r_\infty$ and $\rho_0$ are related as 
$r_\infty ^2 = 4 \left( \rho _0 ^2 + 2 M \rho _0 + Q^2 \right)$. 
The coordinates run the ranges of 
$- \infty < t < \infty ,~ 0 < \rho < \infty ,~ 0 \leq \theta \leq \pi $, $- \pi \leq \phi \leq \pi $ 
and $0 \leq \psi \leq 4 \pi $. 
The squashed Kaluza-Klein black hole solution is asymptotically locally flat, i.e., 
the metric asymptotes to a twisted constant S$^1$ fiber bundle over the four-dimensional Minkowski spacetime. 
The parameters $M ,~Q$ and $r_\infty$ denote the Komar mass, the charge of the black hole and 
the size of the compactified extra dimension at infinity, respectively.

In this paper, to avoid the existence of naked singularities on and outside the horizon, 
we restrict ourselves to the ranges of parameters such that
\begin{align}
 M \geq Q > 0 , \quad r_\infty > 0 ,   
\label{parameters}
\end{align}
with the relation  
\begin{align}
 \rho_0 = \frac{\sqrt{r_\infty ^2 + 4 \left( M^2 - Q^2 \right)} - 2 M}{2} .  
\label{rho0}
\end{align}
In these parameters, the outer and the inner horizons are located at 
$\rho = M + \sqrt{M^2 - Q^2}$ and $\rho = M - \sqrt{M^2 - Q^2}$,  
respectively.  
The parameter $\rho_0$ gives the typical scale of transition from five dimensions to effective four dimensions  
\cite{Matsuno:2011ca}.   
In the limit $\rho_0 \to 0$, equivalently $r_\infty \to 2 Q$, 
we obtain the metric \eqref{met} with $K = 1$ which 
represents the four-dimensional Reissner-Nordstr\"{o}m black hole with a twisted constant S$^1$ fiber.  
We expect the appearance of the higher-dimensional corrections, 
which are related to the parameter $\rho_0$, 
to the photon motions and the deflection angle of photon of four-dimensional relativity.

\section{Photon motions in a plasma medium} 
\label{sec:photonmotions}

We consider motions of photons in the five-dimensional squashed Kaluza-Klein black hole spacetime 
in the presence of an unmagnetized cold plasma medium.  
The Hamiltonian for the photon in the metric \eqref{met} is
\begin{align}
 \mathcal H &= \frac{1}{2} \left( g^{\mu \nu} p_\mu p_\nu + \omega_e^2 \right) 
\notag \\
&= \frac{1}{2} \left( -\frac{p_t ^2}{F} +\frac{F }{K^2} p_\rho ^2 +\frac{p_\theta ^2}{\rho^2 K^2} 
+\frac{ \left( p_\phi - p_\psi \cos \theta \right) ^2}{\rho^2 K^2 \sin ^2 \theta} 
+\frac{4 K^2}{r_\infty ^2} p_\psi ^2 + \omega_e^2 \right)  ,
\end{align}
with $\mathcal H = 0$,   
where $p_\mu$ are the canonical momenta conjugate to the coordinates $x^\mu$ and 
$\omega_e$ is the electron plasma frequency defined by 
$\omega_e ^2 = e^2 N_e / (\epsilon _0 m_e)$,   
where $e ,~ N_e ,~ m_e$ and $\epsilon _0$ are the charge, the number density and the mass of the electron 
in the plasma, and the vacuum permittivity, respectively 
\cite{Perlick(2000), Tsupko:2013cqa, Perlick:2015vta, Crisnejo:2018uyn}.  
Note that we ignore the self-gravitation of the plasma.  
In this paper, we consider a homogeneous plasma with the electron number density $N_e =$ const and 
the positive refractive index $n = \sqrt{1 - \omega_e ^2 F / \omega_\infty ^2}$. 
From the Hamilton's equations, we can obtain three constants of motion as 
\begin{align}
 \omega_\infty := -p_t ,  \quad  L := p_\phi , \quad  \text{and}  \quad  p_\psi ,
\end{align}
where $\omega_\infty$ is the photon frequency measured by an observer at infinity, 
$L$ and $p_\psi$ are angular momenta of the photon in the $\phi$ and the $\psi$ direction, respectively.

Here, we assume that the photon has no momentum in the direction of the extra dimension, i.e., 
$p_\psi = 0$.\footnote{
Since the size of compactified dimension $\psi$ is very small, 
it would be expected that the momenta of massive and massless particles 
conjugate to $\psi$ are hardly excited \cite{Ishihara:1985zm}.  
} 
Then the effective Hamiltonian for the photon is
\begin{align}
 \mathcal H_\text{eff}  
= \frac{1}{2} \left( -\frac{p_t ^2}{F} +\frac{F }{K^2} p_\rho ^2 +\frac{p_\theta ^2}{\rho^2 K^2} 
+\frac{p_\phi ^2}{\rho^2 K^2 \sin ^2 \theta} + \omega_e^2 \right)  ,
\label{effHamiltonian} 
\end{align}
where $\omega_e =$ const and $\mathcal H_\text{eff} = 0$.   
We see that this effective Hamiltonian has the same form 
in the case of the four-dimensional spherically symmetric spacetime 
filled with a homogeneous plasma  
where the plasma frequency $\omega_e$ acts like an effective mass for a photon  
\cite{BisnovatyiKogan:2008yg, BisnovatyiKogan:2010ar}.     
Then we can concentrate on orbits with $\theta = \pi /2$ and $p_\theta = 0$ 
on the assumption of $p_\psi = 0$. 
The Hamilton's equations in these conditions are given by 
\begin{align}
 & \frac{d t}{d \lambda} = \frac{\omega _\infty}{F} ,
\label{Hamiltoneq1}
\\
 & \frac{d \rho}{d \lambda} = \frac{F }{K^2} p_\rho  ,
\label{Hamiltoneq2}
\\
 & \frac{d \phi}{d \lambda} = \frac{L }{\rho ^2 K^2} ,
\label{Hamiltoneq3}
\end{align}
where $\lambda $ is the curve parameter along the photon trajectory. 
The photon frequency satisfies the condition $\omega_\infty  > \omega_e $ 
for the propagation of the photon through the plasma in the squashed Kaluza-Klein spacetime \eqref{met}.

\begin{figure}[!htbp]
\begin{center}
\includegraphics[scale=0.57]{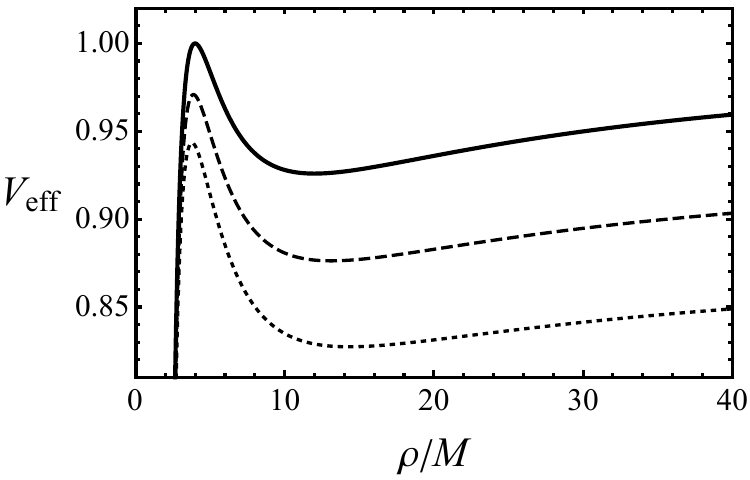} \qquad
\includegraphics[scale=0.57]{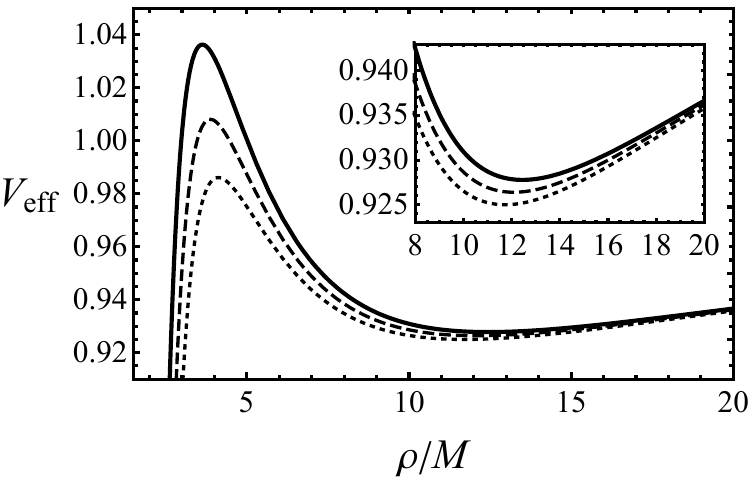}
\caption{
Effective potentials \eqref{effpot} 
in various $\omega_e$ 
for $L / M = 4 ,~ r_\infty / M = 0.1 ,~ Q / M = 0.03$ and $\rho_0 / M \simeq 0.0008$ (left panel). 
$\omega_e = 0.94$ (dotted curve), $\omega_e = 0.97$ (dashed curve) and $\omega_e = 1$ (solid curve). 
The same ones in various $Q / M$ for $L / M = 4 ,~ r_\infty / M = 1$ and $\omega_e = 1$ (right panel). 
$Q = 0$ ($\rho_0 / M \simeq 0.1$, dotted curve), $Q / M = 0.35$ ($\rho_0 / M \simeq 0.06$, dashed curve) 
and $Q / M = 0.5$ ($\rho_0 = 0$, solid curve). 
}
\label{fig:effpot}
\end{center}
\end{figure}

Substituting Eq.~\eqref{Hamiltoneq2} into the Hamiltonian \eqref{effHamiltonian}, 
we obtain the energy conservation equation under the above conditions:  
\begin{align}
 \left( 1 +\frac{\rho_0}{\rho} \right) \left( \frac{d \rho}{d \lambda} \right)^2 
+ V_\text{eff} = \omega_\infty^2 ,
\label{energy}
\end{align}
where the effective potential is given by
\begin{align}
 V_\text{eff} = \left( 1-\frac{2M}{\rho} +\frac{Q^2}{\rho^2} \right) 
\left( \omega_e ^2 +\frac{L^2}{\rho \left( \rho + \rho_0 \right)} \right) . 
\label{effpot}
\end{align}
Typical shapes of the potential $V_\text{eff}$ are shown in Fig.~\ref{fig:effpot}.  
We see that there is a stable circular orbit of a photon at the local minimum of the effective potential.  
Then the squashed Kaluza-Klein black holes in a homogeneous plasma medium, 
where a photon can be stably bounded around the black hole,  
make a remarkable contrast with the higher-dimensional asymptotically flat black holes in such medium, 
which have no stable bound state of photon \cite{Belhaj:2020rdb}.\footnote{
There exist stable bound orbits of massless particles with nonvanishing angular momenta 
in two independent angular directions around an asymptotically flat black ring \cite{Igata:2013be}.} 
From the left panel of Fig.~\ref{fig:effpot},  
we find that the stable circular orbit radius decreases and the unstable one increases 
with increasing $\omega _e$ for fixed $L / M ,~ r_\infty / M$ and $Q / M$. 
From the right panel of Fig.~\ref{fig:effpot},  
we observe that the stable circular orbit radius increases and the unstable one decreases 
with increasing $Q / M$ for fixed $L / M ,~ r_\infty / M$ and $\omega _e$. 
Then the presence of a homogeneous plasma increases the radius of critical photon orbits 
around the Kaluza-Klein black hole similar to the case of the four-dimensional black hole 
surrounded by such plasma \cite{BisnovatyiKogan:2010ar}.    

From $d V_\text{eff} /d\rho = 0$ and $V_\text{eff} = \omega_\infty ^2$,  
we have $\omega_\infty$ and $L$ 
in circular motions with $\rho = R =$ const, $p_\psi = 0$ and $\theta = \pi /2$ as  
\begin{align}
 & \omega_\infty ^2 = \frac{\omega _e^2 \left(\rho _0+2 R\right)
   \left(R^2 -2 M R+Q^2\right)^2}{R^2 \left[ R
   \left(\rho _0 (R-4 M)+2 R (R-3 M)\right)+Q^2
   \left(3 \rho _0+4 R\right) \right]} ,
\label{circularomega}
\\ 
 & L^2 = \frac{2 \omega _e^2 R \left(\rho _0+R\right)^2
   \left(M R-Q^2\right)}{R \left(\rho _0 (R-4
   M)+2 R (R-3 M)\right)+Q^2 \left(3 \rho _0+4
   R\right)} .  
\label{circularL}
\end{align}
Using Eqs.~\eqref{Hamiltoneq1}, \eqref{Hamiltoneq3}, \eqref{circularomega} and \eqref{circularL},  
the period of a circular orbit is given by  
\begin{align}
 T = 2 \pi \frac{d t}{d \phi }
= \sqrt 2 \pi \sqrt{\frac{R^3 (2 R + \rho _0 )}{M R - Q^2}} .  
\end{align}
We see that the orbital period is modified by the extra dimension and the Maxwell field.  
In the large $R$ limit, we have 
\begin{align}
 T \to 2 \pi \sqrt{\frac{R^3 }{M }} . 
\end{align}
This means Kepler's third law.

In the absence of the plasma $\omega _e = 0$, 
Eq.~\eqref{energy} with the effective potential \eqref{effpot} describes null geodesics 
in the five-dimensional charged static squashed Kaluza-Klein spacetime \cite{Sadeghi:2012bj}.  
In this case, an unstable circular orbit of the photon exists at a photonsphere radius. 
From the denominator of the angular momentum \eqref{circularL}, 
we obtain the photonsphere radius in the metric \eqref{met} as  
\begin{align}
 R = & M -\frac{\rho _0}{6} +
\frac{\sqrt{\left(6 M+\rho _0\right)^2-24 Q^2} }{3} 
\notag \\ 
& \times 
\cos \left( \frac{1}{3} \cos ^{-1}\left( \frac{216 M \left( M^2-Q^2 \right) 
+ 18 \rho _0 \left(6 M^2-7 Q^2 -M\rho_0 \right) -\rho _0^3}
{\left(\left(6 M+\rho _0\right)^2-24 Q^2\right)^{3/2}} \right) \right).   
\label{photonsphere}
\end{align} 
In a weak-field limit, 
the photonsphere radius \eqref{photonsphere} becomes 
\begin{align}
 R \simeq 3M \left( 1 + \frac{M \rho_0 - 4 Q^2}{18 M^2} \right) . 
\end{align}
The second term in the right-hand side is the corrections  
by the extra dimension and the Maxwell field.

In the limit $\rho_0 \to 0$, 
the metric \eqref{met} locally has the geometry of the Reissner-Nordstr\"{o}m black string and 
Eq.~\eqref{energy} reduces to the energy conservation equation 
of the photon moving around the four-dimensional Reissner-Nordstr\"{o}m black hole 
surrounded by a homogeneous plasma \cite{Babar:2020txt}.  
In such limit, the energy conservation equation \eqref{energy}  
with the replacements $\omega_e \to 1 ,~ \omega_\infty \to E$ describes   
the geodesic motions of massive test particles with the energy $E$ 
measured by an observer at infinity in the four-dimensional Reissner-Nordstr\"{o}m spacetime \cite{Pugliese:2010ps}.   
This is consistent with a correspondence between motions of the photon with the frequency $\omega_\infty$ 
in an unmagnetized cold homogeneous plasma characterized by the frequency $\omega_e$ 
in a given four-dimensional spacetime 
and timelike geodesics of the massive test particle on the same background   
\cite{Kulsrud:1991jt, BisnovatyiKogan:2010ar}.  
In the limit $Q \to 0$, Eq.~\eqref{energy} and the effective potential \eqref{effpot} 
with the replacements $\omega_e \to 1 ,~ \omega_\infty \to E$ 
describe timelike geodesics of the massive particles with the energy $E$  
in the five-dimensional static squashed Kaluza-Klein spacetime \cite{Matsuno:2009nz}.  
Then some results for photons in a homogeneous plasma shown in the present paper 
would be applied to the geodesic motions and the deflection angles of neutral massive particles 
with $p_\psi = 0$ and $\theta = \pi /2$ 
in the five-dimensional charged static squashed Kaluza-Klein spacetime \eqref{met}.

\section{Deflection angle of photon in a plasma medium}
\label{sec:defangs}

We consider light propagation around the squashed Kaluza-Klein black hole \eqref{met} 
in a homogeneous plasma medium by direct integrations of the photon orbit equation 
with the conditions $p_\psi = 0$ and $\theta = \pi / 2$.    
Substituting Eq.~\eqref{Hamiltoneq3} into Eq.~\eqref{energy} 
and introducing a new coordinate $u = \rho^{-1}$, 
the energy conservation equation becomes 
\begin{align}
 \left( \frac{d u}{d \phi} \right)^2 = \frac{1}{b^2} + \frac{2 \sigma^2 M +\rho_0}{b^2} u 
- \left( 1 + \frac{\sigma^2 (Q^2 - 2 M \rho_0)}{b^2} \right) u^2 
+ \left( 2M -  \frac{\sigma^2 Q^2 \rho_0}{b^2}  \right) u^3 - Q^2 u^4 ,
\label{energy2} 
\end{align}
with 
\begin{align}
 & b^2 := \frac{L^2 }{\omega_\infty ^2 - \omega_e ^2} , 
\label{bandalpha0} 
\\ 
 & \sigma^2 := \frac{\omega_e ^2 }{\omega_\infty ^2 - \omega_e ^2} . 
\label{bandalpha}
\end{align}
Taking the derivative of Eq.~\eqref{energy2} with respect to the coordinate $\phi$, 
we obtain the photon orbit equation 
as 
\begin{align}
 \frac{d^2 u}{d \phi ^2}  = \frac{2 \sigma^2 M +\rho_0}{2 b^2} 
- \left( 1 + \frac{\sigma^2 (Q^2 - 2 M \rho_0)}{b^2} \right) u
+ 3 \left( M -  \frac{\sigma^2 Q^2 \rho_0}{2 b^2}  \right) u^2 - 2 Q^2 u^3 . 
\label{energy3} 
\end{align}
To solve Eq.~\eqref{energy3},  
we assume that the solution $u$ can be expressed in powers of parameters as 
\begin{align}
 u =& u_0 + \frac{M}{b} u_1 + \frac{M^2}{b^2} u_2 + \frac{Q^2}{b^2} u_3 
+ \frac{\rho_0}{b} u_4 + \frac{\rho_0^2}{b^2} u_5 + \frac{M \rho_0}{b^2} u_6 
\notag \\
& + O\left( M^3 ,~ \rho_0 ^3 ,~ M^2 \rho_0 ,~ M Q ^2 ,~ M \rho_0 ^2 ,~ Q^2 \rho_0 \right) ,
\label{u_sol_0}
\end{align}
where $u_i$ ($i = 0 ,~ 1 , \dots ,~ 6$) are functions of $\phi$.  
Substituting the ansatz \eqref{u_sol_0} into Eq.~\eqref{energy3} 
and solving the differential equations of $u_i$, we obtain the orbit of the photon as  
\begin{align}
 u \left( \phi \right) = &\frac{\cos \phi }{b} + \frac{\left(2 \sigma ^2+3\right) M-M \cos (2 \phi )+\rho _0}{2 b^2} 
\notag \\
& + \frac{1}{16 b^3} 
\left[ 
\left(\left(8 \sigma ^2 \left(\sigma ^2+6\right) +37 \right) M^2
+24 M \rho _0 \left(\sigma ^2+1\right) +2 \rho _0^2-\left(8 \sigma ^2+9\right) Q^2 \right) \cos \phi 
\right.
\notag \\
& \quad \left. 
+ \left(3 M^2+Q^2\right) \cos (3 \phi ) 
+ 4 \left(3 M^2 \left(4 \sigma^2+5\right) + \left(2 \sigma ^2+3\right) \left(2 M \rho _0 -Q^2\right) \right) \phi \sin \phi 
\right] 
\notag \\ & 
+ O\left( M^3 ,~ \rho_0 ^3 ,~ M^2 \rho_0 ,~ M Q ^2 ,~ M \rho_0 ^2 ,~ Q^2 \rho_0 \right) , 
\label{u_sol}
\end{align}
where $b$ is the impact parameter which represents the minimum value of 
$\rho$-coordinate for the undeflected light ray, i.e., $M = Q = \rho_0 = 0$.  
We note that the integration constants are chosen such that 
Eq.~\eqref{u_sol} has a symmetry $u (\phi ) = u (- \phi)$  
and satisfies the energy conservation equation \eqref{energy2}    
up to the second order in the parameters $M ,~ Q$ and $\rho_0$.
By taking some limits, we obtain particle trajectories in some four-dimensional spacetimes. 
When $\rho_0 = 0 ,~ \omega_e = 0$, 
Eq.~\eqref{u_sol} represents the orbit of the photon in the braneworld black hole spacetime 
with the tidal charge $q$ which reduces to the four-dimensional Reissner-Nordstr\"{o}m spacetime 
in the limit $q \to Q^2$ \cite{Gergely:2009xg}.  
When $\rho_0 = 0 ,~ Q = 0$, 
Eq.~\eqref{u_sol} with the replacements 
$\phi \to \phi - \pi /2 ,~ \sigma^2 \to \left( 1 - v^2 \right) / v^2$ 
represents the orbit of the neutral massive particle with the velocity $v$ measured by an observer at infinity 
in the four-dimensional Schwarzschild spacetime \cite{Crisnejo:2018uyn}.

We consider the photon which comes from far away at the distant past, 
$\phi = - \pi /2 - \delta \phi / 2$, 
and is deflected by the black hole then travels towards far away at the distant future, 
$\phi = \pi /2 + \delta \phi / 2$, where $\delta \phi $ is a deflection angle. 
Since Eq.~\eqref{u_sol} has a symmetry $u (\phi ) = u (- \phi)$,  
we solve $u (\pi /2 + \delta \phi / 2 ) = 0$ up to the first order in $\delta \phi $. 
Then we obtain the deflection angle of photon in a weak-field limit as 
\begin{align}
\delta \phi 
= & \frac{2 M}{b} \left( 1 + \frac{\rho_0 }{2 M} + \frac{1}{1-\omega_e ^2 / \omega_\infty ^2} \right) 
\notag \\
& + \frac{3 \pi M^2}{4 b^2} \left( 1 + \frac{4}{1-\omega_e ^2 / \omega_\infty ^2} 
+ \frac{2 M \rho_0 - Q^2}{3 M^2} \left( 1 + \frac{2}{1-\omega_e ^2 / \omega_\infty ^2} \right) \right) 
+ O\left( b^{-3} \right) ,
\label{defang1-1}
\end{align}
where 
the parameter $\rho_0$ is given by Eq.~\eqref{rho0} and 
we use Eq.~\eqref{bandalpha} to represent the deflection angle 
in terms of the frequency ratio $\omega_e / \omega_\infty$.  
The deflection angle in terms of the distance of closest approach is shown 
in the Appendix.

\begin{figure}[!htbp]
\begin{center}
\includegraphics[scale=0.57]{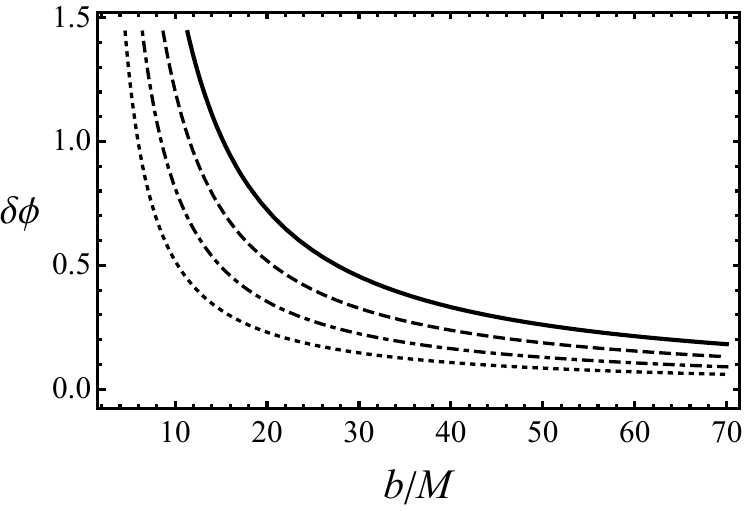} \qquad 
\includegraphics[scale=0.57]{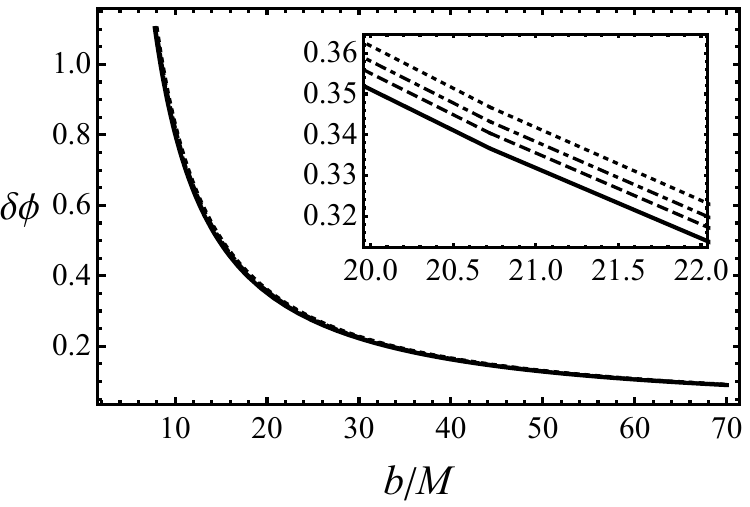}
\caption{
Deflection angles \eqref{defang1-1}  
in various $\omega_e / \omega_\infty $ for $r_\infty / M = 0.1 ,~ Q / M = 0.03$ and 
$\rho _0 / M \simeq 0.0008$ (left panel). 
$\omega_e = 0$ (dotted curve), $\omega_e ^2 / \omega_\infty ^2 = 0.5$ (dot-dashed curve), 
$\omega_e ^2/ \omega_\infty ^2 = 0.7$ (dashed curve) and $\omega_e ^2/ \omega_\infty ^2= 0.8$ (solid curve). 
The same ones in various $Q / M$ for $r_\infty / M = 1$ and $\omega_e ^2/ \omega_\infty ^2 = 0.5$ (right panel). 
$Q = 0$ ($\rho_0 / M \simeq 0.1$, dotted curve), $Q / M = 0.3$ ($\rho_0 / M \simeq 0.08$, dot-dashed curve), 
$Q / M = 0.4$ ($\rho_0 / M \simeq 0.04$, dashed curve) and $Q / M = 0.5$ ($\rho_0 = 0$, solid curve). 
}
\label{fig:defang}
\end{center}
\end{figure}

We see that the gravitational deflection angle \eqref{defang1-1} depends upon the photon frequency $\omega_\infty$ 
and is modified by the squashed Kaluza-Klein geometry, the Maxwell field and the homogeneous plasma 
through the extra dimension size $r_\infty$, the black hole charge $Q$ 
and the plasma frequency $\omega_e$, respectively.  
We find that the deflection angle decreases with increasing $b / M$ 
for fixed $r_\infty / M ,~ Q / M$ and $\omega_e / \omega_\infty$,      
while it increases with increasing $r_\infty / M$ 
for fixed $b / M ,~ Q / M$ and $\omega_e / \omega_\infty$.  
We show the behaviors of the deflection angle $\delta \phi$ versus $b / M$ in Fig.~\ref{fig:defang}.  
From the left panel of Fig.~\ref{fig:defang},  
we see that $\delta \phi$ increases with increasing $\omega_e / \omega_\infty$ 
for fixed $b / M ,~ r_\infty / M$ and $Q / M$. 
Then the presence of plasma changes the deflection angle with the difference 
from the case of no plasma, $\omega_e = 0$ or $\omega_e / \omega_\infty \ll 1$,   
being strongest for photons of smaller frequency or longer wavelength 
as $\omega_\infty$ approaches $\omega_e$.   
From the right panel of Fig.~\ref{fig:defang},  
we observe that $\delta \phi$ decreases with increasing $Q / M$ 
for fixed $b / M ,~ r_\infty / M$ and $\omega_e / \omega_\infty $.  
Then the effect of the difference in gravitational deflection angles is significant for 
smaller charges of the black hole, larger sizes of the extra dimension and 
larger ratios between the plasma and the photon frequencies.   

By taking some limits in Eqs.~\eqref{defang1-1} and \eqref{defang2},   
we obtain second-order deflection angles in some four-dimensional spacetimes.  
First, when $\rho_0 = 0$, equivalently $r_\infty = 2 Q$, 
Eqs.~\eqref{defang1-1} and \eqref{defang2} 
with the replacement $\omega_e ^2/ \omega_\infty ^2 \to 1 - v ^2$ 
coincide with the deflection angles of neutral massive particles 
with the velocities $v$ measured by an observer at infinity 
in the four-dimensional Reissner-Nordstr\"{o}m spacetime \cite{Pang:2018jpm}.  
Second, when $r_\infty = 2 Q ,~ \omega_e = 0$ or $\omega_e / \omega_\infty \ll 1$, 
Eqs.~\eqref{defang1-1} and \eqref{defang2} represent the deflection angles of photons 
in the four-dimensional Reissner-Nordstr\"{o}m spacetime \cite{Briet:2008mz}.  
Third, when $r_\infty = 2 Q$, then taking the limit $Q \to 0$,  
Eqs.~\eqref{defang1-1} and \eqref{defang2} represent the deflection angles of photons 
in a homogeneous plasma medium in the four-dimensional Schwarzschild spacetime \cite{Crisnejo:2018uyn}.  
Lastly, when $r_\infty = 2 Q ,~ \omega_e = 0$ or $\omega_e / \omega_\infty \ll 1$,  
then taking the limit $Q \to 0$,  
Eqs.~\eqref{defang1-1} and \eqref{defang2} represent the deflection angles of photons 
in the four-dimensional Schwarzschild spacetime \cite{Epstein:1980dw, Fischbach:1980su}.

We consider the angular positions of the images determined by the Einstein rings 
as one of the observational effects in a gravitational lensing \cite{BisnovatyiKogan:2010ar}. 
Using the deflection angle \eqref{defang1-1} and the lens equation 
in the case of a perfect alignment of the source, the lens and the observer,  
we obtain the Einstein ring $\theta_\text{KK}$ in the squashed Kaluza-Klein spacetime 
in the presence of the plasma as 
\begin{align}
\theta_\text{KK}
= \sqrt{ 2 M \left( 1 + \frac{\rho_0 }{2 M} + \frac{1}{1-\omega_e ^2 / \omega_\infty ^2} \right) 
\frac{D_{LS}}{D_L D_S} } ,
\end{align}
where $D_{LS}$ is the distance between the lens and the source, 
$D_{L}$ is the distance between the observer and the lens, 
$D_{S}$ is the distance between the observer and the source, 
and we use the relation $\delta \phi \sim b^{-1}$ 
in order to express the small angle $\theta_\text{KK} = b / D_{L}$ in the weak-field approximation  
\cite{Crisnejo:2018uyn, Turimov:2018ttf}.  
Then the relative change in the position of the Einstein rings 
between $\theta_\text{KK}$ and $\theta_\text{Sch} := \sqrt{4 M D_{LS} / (D_L D_S)}$ 
in the four-dimensional Schwarzschild spacetime is given by 
\begin{align}
\frac{\Delta \theta_\text{KK} }{\theta_\text{Sch} } 
:= \frac{\theta_\text{KK} - \theta_\text{Sch} }{\theta_\text{Sch} } 
= \sqrt{ \frac{1}{2} \left( 1 + \frac{\rho_0 }{2 M} + \frac{1}{1-\omega_e ^2 / \omega_\infty ^2} \right)} -1 .
\end{align}
For $\omega _e /\omega _\infty \ll 1 ,~ Q / M \ll 1 ,~ r_\infty / M \ll 1$, 
we obtain the relative change in the position of the images as 
\begin{align}
\frac{\Delta \theta_\text{KK} }{\theta_\text{Sch} } = \delta _1 - \delta _2 + \delta _3 ,
\end{align}
with 
\begin{align}
\delta _1 := \frac{\omega _e ^2}{4 \omega _\infty ^2} , \quad
\delta _2 := \frac{Q^2}{16 M^2} , \quad
\delta _3 := \frac{r_\infty ^2}{64 M^2} .
\end{align}
First, we consider the correction $\delta _1$ by the plasma. 
We see that  the correction $\delta _1$ coincides with that 
in a homogeneous plasma medium in the four-dimensional Schwarzschild spacetime 
\cite{BisnovatyiKogan:2010ar}.  
Then, for the photon frequency $\omega _\infty / (2 \pi) \simeq 3 \times 10^8$ Hz and 
the electron number density $N_e \simeq 5 \times 10^{10}$ m$^{-3}$, 
we can estimate that $\delta _1$ is of order $10^{-5}$ 
with a value of $\theta _\text{Sch} \simeq 1$ arcsec, 
which would be detected in near future observations 
\cite{BisnovatyiKogan:2010ar, Crisnejo:2018uyn}.  
Second, we consider the correction $\delta _2$ by the black hole charge.  
For example, a theoretical uppermost limit on the charge of the supermassive black hole Sgr A$^*$ 
with the mass $\sim 4 \times 10^6 M_\odot$ is $\lesssim 7 \times 10^{26}$ C 
by considering an extremal charged black hole, 
while an observational upper limit is $\lesssim 3 \times 10^8$ C 
by the mass difference between the proton and the electron in a plasma around the Sgr A$^*$, 
where $M_\odot \simeq 2 \times 10^{30}$ kg is the mass of the Sun \cite{Zajacek:2018ycb}. 
Using these constraints, we can estimate that $\delta _2 \lesssim 0.06$ for the theoretical limit 
and $\delta _2 \lesssim 10 ^{-38}$ for the observational limit.  
Then we see that, though a black hole charge could modify a gravitational lensing theoretically,    
the correction $\delta _2$ for a supermassive black hole 
would not appear to be relevant for present and near future observations.   
Lastly, we consider the correction $\delta _3$ by the compactified extra dimension.  
If the size of the extra dimension is $r_\infty \simeq 0.1$ mm \cite{Matsuno:2009nz},   
we can estimate that $\delta _3 \simeq 10^{-30}$ for the Sgr A$^*$,  
$\delta _3 \simeq 10^{-18}$ for a stellar black hole with the mass $\sim 10 M_\odot$ \cite{Turimov:2018ttf},   
and $\delta _3 \simeq 10^{-5}$ for a primordial black hole  
with the mass of the Earth $\sim 3 \times 10^{-6} M_\odot$ \cite{Sasaki:2018dmp}.    
Then we see that it would be difficult to detect $\delta _3$ for supermassive and stellar black holes 
in present and near future observations, 
while it would be challenging to detect $\delta _3$ for primordial black holes 
in future observations of a gravitational lensing 
\cite{Sasaki:2018dmp, Keeton:2006di, Naderi:2017ite, Carr:2020gox}.

\section{Summary and discussion}
\label{sec:summary}

We consider motions of photons around a spherical compact object 
in an unmagnetized cold homogeneous plasma medium.   
We assume that the five-dimensional charged static squashed Kaluza-Klein black hole metric   
describes the geometry of the region outside the compact object 
and a photon has no momentum in the direction of the extra dimension.         
We show that 
the five-dimensional squashed Kaluza-Klein spacetime in the presence of a homogeneous plasma 
admits stable circular orbits of photons 
similar to the four-dimensional spherically symmetric black holes surrounded by such plasma.         
We solve the photon orbit equation in the plasma medium in the squashed Kaluza-Klein spacetime  
and derive the deflection angle of photon in a weak-field limit 
with corrections by the extra dimension, the Maxwell field and the plasma.   
Some known deflection angles in four-dimensional spacetimes 
are obtained by taking limits in our deflection angle of photon. 
We see that, for fixed values of the photon frequency and the black hole mass, 
the deflection angle of photon  
increases with increasing the extra dimension size and the plasma frequency, 
while decreases with increasing the impact parameter and the black hole charge.  
The variations of these parameters provide specific signatures on the optical features 
of the squashed Kaluza-Klein black hole solutions in the plasma medium   
which would open the possibility of testing such higher-dimensional models 
by using astronomical and astrophysical observations.  
We consider the difference between angular positions of images 
in the squashed Kaluza-Klein spacetime 
and in the four-dimensional Schwarzschild spacetime,  
and estimate its corrections by the plasma, the black hole charge and the extra dimension.  
We see that the correction by the plasma would be detectable in near future observations, 
while the other two corrections for supermassive and stellar black holes 
would not appear to be relevant for present and near future observations. 
However, it would be expected that the correction by the extra dimension 
might be detected in future observations of a gravitational lensing by primordial black holes. 
If a precise observation of a gravitational lensing by an astrophysical black hole 
agrees with the expected value of general relativity, 
it requires a rigorous upper limit of the size of the extra dimension, 
or it excludes the squashed Kaluza-Klein metric for describing the geometry around such a black hole.

We note that the exterior spacetimes of standard general relativistic spherical compact objects 
are described by the Schwarzschild metric.   
However, in higher-dimensional spacetime models with Kaluza-Klein structures,   
the Schwarzschild metric is no longer the exterior metric of a static compact object.    
Even if we impose asymptotic flatness to the four-dimensional part of the spacetime, 
there are various possibilities of fiber bundle structures of the extra dimensions 
as the fiber over the four-dimensional base spacetime. 
For example, the direct product of four-dimensional Schwarzschild spacetime 
with a small S$^1$ is a possible metric to describe the exterior of the compact object.   
In this case, no higher-dimensional correction of light deflection appears 
without a momentum of the photon in the direction of the extra dimension.   
In contrast, it is interesting that the correction exists 
even if the photon moves along the four-dimensional spacetime in the squashed Kaluza-Klein geometry.

Since the asymptotically Kaluza-Klein structure, the Maxwell field and the homogeneous plasma 
modify the unstable circular orbits of photons in the four-dimensional Schwarzschild spacetime, 
shadows of black holes would be influenced by such corrections.       
Moreover, generalizations of the present study to light deflections in inhomogeneous plasma media  
\cite{BisnovatyiKogan:2010ar, Rogers:2015dla, Crisnejo:2018ppm} 
in another class of Kaluza-Klein type metrics 
\cite{Gross:1983hb, Liu:1997fg} 
would be interesting.  
We leave the analysis of these topics for the future.

\section*{Acknowledgments}
The author would like to thank Dr. Hideki Ishihara, Dr. Ken-ichi Nakao, Dr. Hirotaka Yoshino 
and colleagues at the theoretical astrophysics and gravity group in Osaka City University 
for valuable suggestions and discussions.  
This work was partly supported by Osaka City University Advanced Mathematical Institute 
(MEXT Joint Usage/Research Center on Mathematics and Theoretical Physics JPMXP0619217849).

\appendix*

\section{Deflection angle of photon in terms of distance of closest approach}
\label{appendix:deflectionangle}

While the deflection angle is given in terms of the impact parameter $b$,  
it would be useful to represent the deflection angle in terms of the distance of closest approach $\rho_\text{min}$. 
Substituting $\phi = 0$ and $u = \rho_\text{min}^{-1}$ into Eq.~\eqref{u_sol}, 
the relation between the distance of closest approach and the impact parameter is given by 
\begin{align}
\frac{1}{\rho_\text{min} } = \frac{1}{b} \left( 1 + \frac{2 M (\sigma^2 + 1) + \rho_0 }{2 b} 
+ \frac{4 (\sigma^2 +1) ((\sigma^2 + 5) M^2 -Q^2) +12 M \rho_0 (\sigma^2 + 1) + \rho_0^2 }{8 b^2} \right) .
\end{align}
Solving this equation for $b^{-1}$ up to the second order in $\rho_\text{min}^{-1}$, we have 
\begin{align}
\frac{1}{b} \simeq \frac{1}{\rho_\text{min} } \left( 1 - \frac{2 M (\sigma^2 + 1) + \rho_0 }{2 \rho_\text{min} } \right) .
\end{align}
Then we obtain the deflection angle in terms of the distance of closest approach as 
\begin{align}
\delta \phi = & 
\frac{4 M}{\rho_\text{min} } \left( 1 + \frac{\sigma^2}{2} + \frac{\rho_0}{4 M} \right) 
+ \frac{15 \pi M^2}{4 \rho_\text{min}^2} 
\left( 1 + \frac{4 \sigma^2}{5} + \frac{(2 \sigma^2 +3) (2 M \rho_0 -Q^2)}{15 M^2} \right) 
\notag \\ & 
- \frac{4 M^2}{\rho_\text{min}^2 } 
\left( 1 + \frac{\sigma^2 (\sigma^2 + 3) }{2} + \frac{\rho_0 \left( 2 M (2 \sigma^2 +3) +\rho_0 \right)}{8 M^2} \right) 
+ O\left( \rho_\text{min}^{-3} \right) .
\label{defang2}
\end{align}

\end{document}